\documentclass{article}

\usepackage{lineno,hyperref}

\usepackage{url}
\usepackage{booktabs}
\usepackage{graphicx}
\usepackage{color}
\usepackage{array}
\usepackage{hyperref}
\usepackage{ragged2e}
\usepackage{enumitem}
\usepackage[T1]{fontenc}
\usepackage{lmodern}
\usepackage{amsmath}

\usepackage[utf8]{inputenc}
\usepackage{csquotes}
\usepackage[super]{nth}
 
\usepackage{xspace}
\usepackage{minibox}

\usepackage{multirow}
\usepackage{tabularx}

\newcommand{\mypara}[1]{\vspace{0.1in}\noindent\textit{#1}}

\newcommand*\rot{\rotatebox{90}}

\newcolumntype{L}[1]{>{\raggedright\let\newline\\\arraybackslash\hspace{0pt}}m{#1}}
\newcolumntype{C}[1]{>{\centering\let\newline\\\arraybackslash\hspace{0pt}}m{#1}}
\newcounter{EKXCommentsCounter}

\begin{document}


\title{Use of Agile Practices in Start-ups}


\author{Eriks Klotins \and Michael Unterkalmsteiner \and Panagiota Chatzipetrou \and Tony Gorschek \and Rafael Prikladnicki \and Nirnaya Tripathi \and Leandro Bento Pompermaier}

\date{}

\maketitle





\begin{abstract}
\emph{Context:} Software start-ups have shown their ability to develop and launch innovative software products and services. Small, motivated teams and uncertain project scope makes start-ups good candidates for adopting Agile practices.

\emph{Objective:} We explore how start-ups use Agile practices and what effects can be associated with the use of those practices.

\emph{Method:} We use a case survey to analyze 84 start-up cases and 56 Agile practices. We apply statistical methods to test for statistically significant associations between the use of Agile practices, team, and product factors.

\emph{Results:} Our results suggest that backlog, version control, refactoring, and user stories are the most frequently reported practices. We identify 22 associations between the use of Agile practices, team, and product factors. The use of Agile practices is associated with effects on source code and overall product quality. Teams' attitudes towards following the best engineering practices are the leading precedents for adopting several Agile practices.
 To explore causal relationships in our findings, we set forth a number of propositions that can be investigated by future research.

\emph{Conclusions:} We conclude that start-ups use Agile practices, however without following any specific methodology. We identify the opportunity for more fine-grained studies into the adoption and effects of individual Agile practices. Start-up practitioners could benefit from Agile practices in terms of better overall quality, tighter control over team performance and resource utilization.

\end{abstract}




\section{Introduction}






Start-ups are important suppliers of innovation, new software products, and services. However, engineering the software in start-ups is a complicated endeavor as the start-up context poses challenges to software engineers~\cite{Giardino2015}. 
As a result of these challenges, most start-ups 
do not survive the first few years of operation and cease to exist before 
delivering any value~\cite{Blank2013,Giardino2014}.

Uncertainty, changing goals, limited human resources, extreme time and resource 
constraints are reported as characteristic to start-ups~\cite{Paternoster2014a, Giardino2015}.

To survive in such a context, start-ups use ad-hoc engineering practices and attempt to tailor agile methods to their needs. However, scaled-down agile methods could be irrelevant and ignore start-up specific challenges~\cite{klotins2019software,Yau2013}.

Giardino et al.~\cite{Unterkalmsteiner} suggest that start-ups adopt practices as a response to some problematic situations and do not consider adopting full agile methodologies, e.g., scrum or XP, at least in early stages.

Pantiuchina et al.~\cite{pantiuchina2017software} make a similar observation and argue that start-ups focus more on speed-related practices, e.g., iterations and frequent releases, than quality-related practices, e.g., unit testing and refactoring.


In this study, we explore the use of Agile practices in start-ups. We focus on identifying the associations between certain Agile practices, product, and team factors. We aim to understand what positive, and potentially adverse effects can be associated with the use of specific practices. We use our results to formulate propositions for further exploration. 

We use a case survey to collect data from 84 start-up cases~\cite{klotins2019progression}. We use statistical methods to analyze 11,088 data points and identify associations between the use of Agile practices and respondents' estimates on various team and product factors. 

We identify 20 statistically significant associations pointing towards potential causes and effects of using Agile practices. We identify that the use of automated tests and continuous integration is associated with positive attitudes towards following best practice. However, the use of planning and control practices are more associated with deficiencies in the attitudes. 

The rest of this paper is structured as follows. In Section 2, we discuss related work. Section 3 covers the research methodology, data collection, and our approach to data analysis. Section 4 presents the results. We answer our research questions and discuss implications for research and practice in Section 5. Section 6 concludes the paper.

\section{Background and Related Work}

\subsection{Software Start-ups}
Software start-ups are small companies created for the purpose of developing 
and bringing an innovative product or service to market, and to benefit
from economies of scale. 

Start-up companies rely on external funding to support their endeavors. In 2015 alone, start-up companies have received investments of 429 billion USD in the US and Europe alone~\cite{PitchBookData2015a,PitchBookData2015}. With an optimistic start-up failure rate of 75\% that constitutes of 322 billion USD of capital potentially wasted on building unsuccessful products. 

Earlier studies show that product engineering challenges and inadequacies in 
applied engineering practices could be linked to start-up 
failures~\cite{Giardino2015,klotins2018software}. To what extent software engineering 
practices are responsible or linked to success rate is very hard to judge. 
However, if improved software engineering practices could increase the 
likelihood of success by only a few percent, it would yield a significant impact 
on capital return.

Some authors, e.g. Sutton~\cite{Sutton2000} and Giardino~\cite{Giardino2014}, point out the unique challenges in start-ups, such as high risk, uncertainty, lack of resources, rapid 
evolution, immature teams, and time pressure among other factors. However, 
start-ups are flexible to adopt new engineering practices, and reactive to keep up with emerging technologies and markets~\cite{Unterkalmsteiner}.
However, our earlier study~\cite{klotins2018software} analyzing the amount of empirical evidence supporting the uniqueness of start-ups found that most start-up characteristics are based on anecdotal evidence. Thus, there could be negligible difference between start-ups and other organizations launching new products to market in terms of software engineering.

\subsection{Agile practices}

Agile software engineering practices originate from the Agile manifesto proposing a shift
from heavyweight, plan-driven engineering towards more lightweight, customer-oriented,
and flexible methodologies~\cite{beck2001manifesto}. Agile methodologies, such as Scrum, XP, and Kanban, 
prescribe specific sets of Agile practices~\cite{Rising2000, jyothi2012effective}. However, in practice, by-the-book methodologies are often tailored with additional practices to address specific concerns~\cite{diebold2014agile,jalali2012global}. Thus, we focus our study on what practices start-ups use, without considering any specific agile methodology.

Small organizations have successfully adopted Agile practices for projects where requirements are uncertain and expected to change~\cite{Chow2008,Misra2012}. 
In theory, Agile practices could be perfect for software start-ups~\cite{Yau2013}. However, successful adoption
of Agile practices requires highly skilled teams and support throughout an organization~\cite{solinski2016prioritizing,Chow2008}.

Earlier work on engineering practices in start-ups suggests that start-ups initially rely on an ad-hoc approach to engineering and adopt agile principles incrementally when a need for more systematic practices arises. The shift is often motivated by excessive technical debt hindering quality and lack of control over the engineering process~\cite{Unterkalmsteiner}. 

We explore associations between 56 Agile practices and aspects of technical
debt to pinpoint relevant practices for prevention and early identification of technical debt.
We use a list and descriptions of Agile practices compiled by Agile Alliance, a non-profit community promoting agile principles~\cite{agilealliance}. To our best knowledge, their website contains the most comprehensive list of Agile practices to date.

In this study we consider the following practices whose definitions can be found at the Agile Alliance's website~\cite{agilealliance}: Card, Conversation, Confirmation (3C's), Acceptance tests, Acceptance Test-Driven Development (ATDD), Automated build, Backlog, Backlog grooming, Behavior Driven Development, Burndown chart, Collective ownership, Continuous deployment, Continuous integration, Class Responsibility Collaborator (CRC) Cards cards, Daily meeting, Definition of Done, Definition of Ready, Exploratory testing, Facilitation, Frequent releases, Given-When-Then, Heartbeat retrospective, Incremental development, INVEST, Iterations, Iterative development, Kanban board, Lead time, Mock objects, Niko-Niko, Pair Programming, Personas, Planning poker, Point estimates, Project charters, Quick design session, Refactoring, Relative estimation, Role-Feature-Reason, Rules of simplicity, Scrum of Scrums, Sign up for tasks, Simple design, Story mapping, Story splitting, Sustainable Pace, Task board, Team, Team room, Test-driven development, Three Questions, Timebox, Ubiquitous language, Unit tests, Usability testing, User stories, Velocity, and Version control.

\subsection{Effects of using Agile practices}

The use of Agile practices is associated with increased product quality and fewer defects compared to plan-driven approaches~\cite{layman2004exploring,ilieva2004analyses}. We analyze associations between use of the Agile practices, product documentation, software architecture, quality of the source code, tests, and the overall product quality. In this paper, we adopt the product view on software quality, recognizing the relationship between internal product characteristics and quality in use~\cite{kitchenham1996software}.





Product documentation comprises of written requirements, architecture documentation, and test cases. Deficiencies in such artifacts are associated with hindered knowledge distribution in the team and with adverse effects on further development and maintenance of the product~\cite{Tom2013}. Note that we analyze if documentation artifacts are understandable and useful without implying any specific format.

Even though the Agile manifesto emphasizes working software over comprehensive documentation, some documentation is essential~\cite{beck2001manifesto}. For example, user stories are one of the key agile tools to document requirements~\cite{lucassen2015forging}. System metaphor is useful to communicate the logical structure of the software to all stakeholders~\cite{khaled2004system}. The use of automated testing in continuous integration and deployment pipelines require formally defined tests~\cite{collins2012software}.

Software architecture denotes how different components, modules, and technologies are combined to compose the product. Symptoms such as outdated components, a need for workarounds and patches point towards deficiencies in the software architecture and the lack of attention to refactoring~\cite{moser2007case,selic2009agile}.

Source code quality is determined by the use of coding standards and refactoring practices~\cite{palomba2014they,Mantyla2003}.
Degrading architecture and poorly organized source code is associated with increased software complexity, difficult maintenance, and product quality issues down the road~\cite{Tom2013}.

We analyze the quality (or lack, thereof) of automated test scripts removing the need to perform manual regression testing with every release of the product. The effort of manual regression testing grows exponentially with the number of features, slowing down release cycles and making defect detection a time consuming and tedious task~\cite{Tom2013}. 

We also examine associations between product quality and the use of Agile practices. With product quality, we understand non-functional aspects of the product, such as performance, scalability, maintainability, security, robustness, and the ability to capture any defects before the product is released to customers~\cite{Tom2013}.


Good communication, teamwork, adequate skills, and a positive attitude towards following the best practices are recognized as essential team factors for project success~\cite{Chow2008}. Agile software engineering practices aim to facilitate communication, empower individuals, and improve teamwork~\cite{dybaa2008empirical}.
We analyze the associations between team characteristics and the use of specific Agile practices.

Attitudes determine the level of apathy or interest in adopting and following the best engineering practices. Skills characterize to what extent individual members of a start-up team possess relevant engineering skills and knowledge. Communication captures to what extent the team can communicate and coordinate the engineering work. Giardino et al.~\cite{Unterkalmsteiner} identify the team as the catalyst for product development in start-ups. Sufficient skills, positive attitudes, and efficient communication are essential for rapid product development in both agile and start-up contexts~\cite{Unterkalmsteiner,Chow2008}. 

Pragmatism characterizes to what extent a team can handle trade-offs between investing in perfected engineering solutions and time-to-market. 
Agile practices advocate for frequent releases and good-enough solutions~\cite{Rising2000}. Such practices help to validate the product features early and gather additional feedback from customers~\cite{klotins2018software}. On the other hand, quick product releases need to be accompanied by frequent refactoring and unit tests to manage technical debt and keep regression defects under control~\cite{Chow2008}. Start-ups often overlook such corrective practices~\cite{Unterkalmsteiner,klotins2018software}.

Sufficient time and resources for product engineering are essential for project success~\cite{Chow2008}. We analyze what Agile practices can be associated with better resource estimation and planning in start-ups. Several authors, e.g., Giardino et al.~\cite{Giardino2014} and Sutton~\cite{Sutton2000} identify resource shortages as one of the critical challenges in start-ups. However, we, in our earlier study identify the lack of adequate resources planning and control practices in early start-ups~\cite{klotins2019progression}.

Process characterizes to what extent product engineering is hindered by unanticipated changes in organizational priorities, goals, and unsystematic changes in the product itself. Lack of organizational support for agile product engineering contributes to project failures~\cite{Chow2008}. On the other hand, Agile practices offer some room for adjusting to unclear and changing objectives~\cite{Misra2012}.


\section{Research methodology}

\subsection{Research aim}

We aim to explore how start-ups use Agile practices and what positive and negative effects can be associated with specific practices.

\subsection{Research questions}

To guide our study, we define the following research
questions (RQ):

\mypara{\textbf{RQ1:}} How are Agile practices used in start-ups?

\mypara{Rationale:} 
With this question, we identify what Agile practices and in what combinations start-ups use. 

\mypara{\textbf{RQ2:}} What are the associations between specific Agile practices and product factors?

\mypara{Rationale:} With this question, we explore the associations between specific Agile practices, quality of documentation, architecture, source code, testing, and overall product quality.

\mypara{\textbf{RQ3:}} What are the associations between specific Agile practices and team factors?

\mypara{Rationale:} With this question, we explore the associations between specific Agile practices, attitudes towards following best engineering practices, pragmatism, communication, skills, resources, engineer process, and teams' productivity.

\subsection{Data collection}

We used a case survey method to collect primary data from start-up 
companies~\cite{klotins2019progression,Larsson1993}. 

The case survey method is based on a questionnaire and is a compromise between a traditional case study and a regular survey~\cite{Klotins2017}. We have designed the questionnaire to collect 
practitioners' experiences in specific start-up cases.

During the questionnaire design phase, we conducted multiple internal and external reviews to ensure that all questions are relevant, clear and that we receive meaningful answers. First, the questions were reviewed in multiple rounds by the first three authors of this paper to refine the scope of the survey and question formulations. Then, with the help of other researchers from the 
Software Start-up Research Network\footnote{The Software Start-up Research 
Network, \url{https://softwarestartups.org/}}, we conducted a workshop to gain external input on the questionnaire. A total of 10 researchers participated and provided their input.

Finally, we piloted the questionnaire with four practitioners from different start-ups. During the pilots, respondents filled in the questionnaire while discussing questions, their answers, and any issues with the first author of this paper. 

As a result of these reviews, we improved the question formulations and removed 
some irrelevant questions. The finalized questionnaire contains 85 
questions in 10 sections. The questionnaire captures 285 variables from 
each start-up case. The full questionnaire is available as supplemental material on-line\footnote{Full questionnaire:~\url{http://eriksklotins.lv/files/GCP questionnaire.pdf}}.

From all the variables, 45 variables focus on capturing the magnitude of 
dimensions, precedents, and outcomes linked to technical debt
The questions capture the respondents' agreement with a statement on a Likert scale: not at all (1), a little (2), somewhat (3), very much (4). 
The values indicate the degree of agreement with a statement. Statements are formulated consistently in a way that lower values indicate less, and higher values indicate more agreement with the statement.

We use a list of 56 Agile practices to capture respondent's answers on what practices they use in their companies~\cite{agilealliance}. The answers are captured in a binary, use or not use, format. In addition to specific practices, we offer an ``I do not know'' and ``other'' options to accommodate for lack of respondents knowledge and to discover other, unlisted, practices.

In addition to questions about software engineering, the questionnaire contains questions inquiring about the engineering context in the start-up and applied software engineering practices. 

The data collection took place between December 1, 2016, and June 15, 2017. The survey was promoted through personal contacts, by attending industry events, and with posts on social media websites. Moreover, we invited other researchers from the Software Start-up Research Network to collaborate on the data collection. This collaboration helped to spread the survey across many geographical locations in Europe, North and South America, and Asia.

\subsection{Data analysis methods}

To analyze the survey responses, we used several techniques. We started by
screening the data and filtering out duplicate cases, responses with few questions answered, or otherwise unusable responses. In the screening, we attempt to be as
inclusive as possible and do not remove any cases based on the provided responses.

Overall, we analyze responses from 84 start-up cases, 132 data-points per each case,  and 11,088 data-points. We use the Chi-Squared test of association to test if the associations between the examined variables are not due to chance. To prevent Type I errors, we used exact tests, specifically, the Monte-Carlo test of statistical significance based on 10,000 sampled tables and assuming $(p < 0.05)$~\cite{hope1968simplified}. 


To examine the strength of associations, we use Cramer's V test. We interpret the test results as suggested by Cohen~\cite{cohan1988statistical}, see Table~\ref{table_cramersv}.
To explore specifics of the association, such as which cases are responsible for this association, we perform post-hoc testing using adjusted residuals. We 
consider an adjusted residual significant if the absolute value is above 1.96 
$(Adj.residual > 1.96)$, as suggested by 
Agresti~\cite{agresti1996introduction}. 

The adjusted residuals drive our analysis on how different groups of start-ups estimate aspects of technical debt. However, due to the exploratory nature of our study, we do not state any hypotheses upfront and drive our analysis with the research questions.

\begin{table}[!t]
\renewcommand{\arraystretch}{1.1}
\caption{Interpretation of Cramer's V test}
\label{table_cramersv}
\centering

\begin{tabular}{C{1in}L{1.5in}}

Cramer's V value & Interpretation   \\
\hline
$\ge 0.1 $ & Weak association       \\
$\ge 0.3 $ & Moderate association   \\
$\ge 0.5 $ & Strong association     \\
\hline
\end{tabular}
\end{table}



\subsection{Validity threats}

In this section, we follow guidelines by Runeson et al.~\cite{Runeson2012} and discuss four types of validity threats and applied countermeasures in the context of our study.

\subsubsection{Construct validity}
Construct validity concerns whether operational measures represent the studied subject~\cite{Runeson2012}. A potential threat is that the statements we use to capture respondent estimates are not capturing the factors. 

To address this threat, we organized a series of workshops with other researchers and potential respondents to ensure that questions are clear, to the point, and to capture the studied phenomenon.

We triangulate each factor by capturing it by 3 - 4 different questions in the questionnaire. 
To avoid biases stemming from respondents preconceived opinions about the effects of agile
practices, we separate questions about the use of the practices and questions inquiring about team and product factors.

To accommodate for the fact that a respondent may not know answers to some of 
the questions, we provide an explicit ``I do not know'' answer option to all 
Likert scale questions. 

\subsubsection{Internal validity}

This type of validity threat addresses causal relationships in the study 
design~\cite{Runeson2012}. 
With this study, we attempt to explore causal relationships between the use of Agile practices and their effects on teams and products. However, due to the nature of our data, we cannot establish causal relationships with confidence. We address this threat by explaining our findings with propositions. The propositions are aimed at providing alternative explanations to our findings and inviting further research.

\subsubsection{External validity}

This type of validity threat concerns to what extent the results could be valid to start-ups outside the study~\cite{Runeson2012}. The study setting for participants was close to real life as possible. That is, the questionnaire was filled in without researcher intervention and in the participant's environment.

The sampling of participants is a concern to external validity. We use convenience sampling to recruit respondents and with the help of other researchers, distributed the survey across several different start-up communities. Demographic information from respondent answers shows that our sample is skewed towards active companies, respondents with little experience in start-ups, young companies, and small development teams of 1-8 engineers. In these aspects, our sample fits the general characteristics of start-ups, see, for example, Giardino et al.~\cite{Giardino2014, Giardino2015} and Klotins et al.~\cite{klotins2019software}. However, there is a 
survivor bias, that is, failed start-ups are under-represented. Thus our results reflect state-of-practice in active start-ups. 

Another threat to external validity stems from case selection. We marketed the questionnaire to start-ups building software-intensive products. However, due to the broad definition of software start-ups (see Giardino et 
al.~\cite{Giardino2014}), it is difficult to differentiate between start-ups and small-medium enterprises. We opted to be as inclusive as possible and to discuss relevant demographic information along with our findings.

\subsubsection{Conclusion validity}

This type of validity threat concerns the possibility of incorrect interpretations arising from flaws in, for example, instrumentation, respondent and researcher personal biases, and external influences~\cite{Runeson2012}. 

To make sure that respondents interpret the questions in the intended way we 
conducted several pilots, workshops and improved the questionnaire 
afterwards. To minimize the risk of systematic errors, the calculations and 
the first and the third author performed statistical analysis 
independently, and findings were discussed among the authors.

To test if the order of appearance of Agile practices affects practitioner responses, we run a Spearman's rank-order correlation test~\cite{daniel1990spearman}. We examine a potential relationship between the order of appearance and the frequency chosen by respondents. The results showed that there is no statistically significant correlation ($p > 0.05$).


\section{Results}

The majority of the surveyed start-ups (63 out of 84, 74\%) are active and had been operating for 1\,-\,5 years (58 out of 84, 69\%).
Start-ups are geographically distributed among Europe (34 out of 86, 40\%), South America (41 
out of 84, 49\%), Asia (7 out of 84), and North America (2 out of 84).

Our sample is about equally distributed in terms of the product development phase. We follow the start-up life-cycle model proposed by 
Crowne~\cite{Crowne2002} and distinguish between inception, stabilization, growth, and maturity phases. In our sample, 16 start-ups have been working on a 
product but have not yet released it to market, 24 teams had released the first 
version and actively develop it further with customer input, 26 start-ups have 
a stable product and they focus on gaining customer base, and another 16 
start-ups have mature products, and they focus on developing variations of their 
products. 

The start-ups in our sample do per-customer customization to some extent: 10 companies (11\%) had specified that they tailor each product 
instance to a specific customer, 30 companies (35\%) do not do per-customer 
customization at all, while 43 start-ups (49\%) occasionally perform product 
customization for an individual customer.

The questionnaire was filled in mostly by start-up founders (64 out of 86, 
76\%) and engineers employed by start-ups (15 out of 86, 18\%). About half of respondents have specified that their area of expertise is software engineering (49 out of 86, 58\%). Others have specified marketing, their respective domain, and business development as their areas of expertise.

The respondents' length of software engineering experience ranges from 6 months to more than 10 years. A large portion of respondents (44 out of 86, 52\%) had 
less than 6 months of experience in working with start-ups at the time when 
they joined their current start-up.

We provide a complete list of studied cases and their demographical information as supplemental material on-line\footnote{The studied cases:~\url{http://eriksklotins.lv/files/GCP_demographics.pdf}}.

 \begin{figure}[ht]
    \centering
    \includegraphics[width=\textwidth]{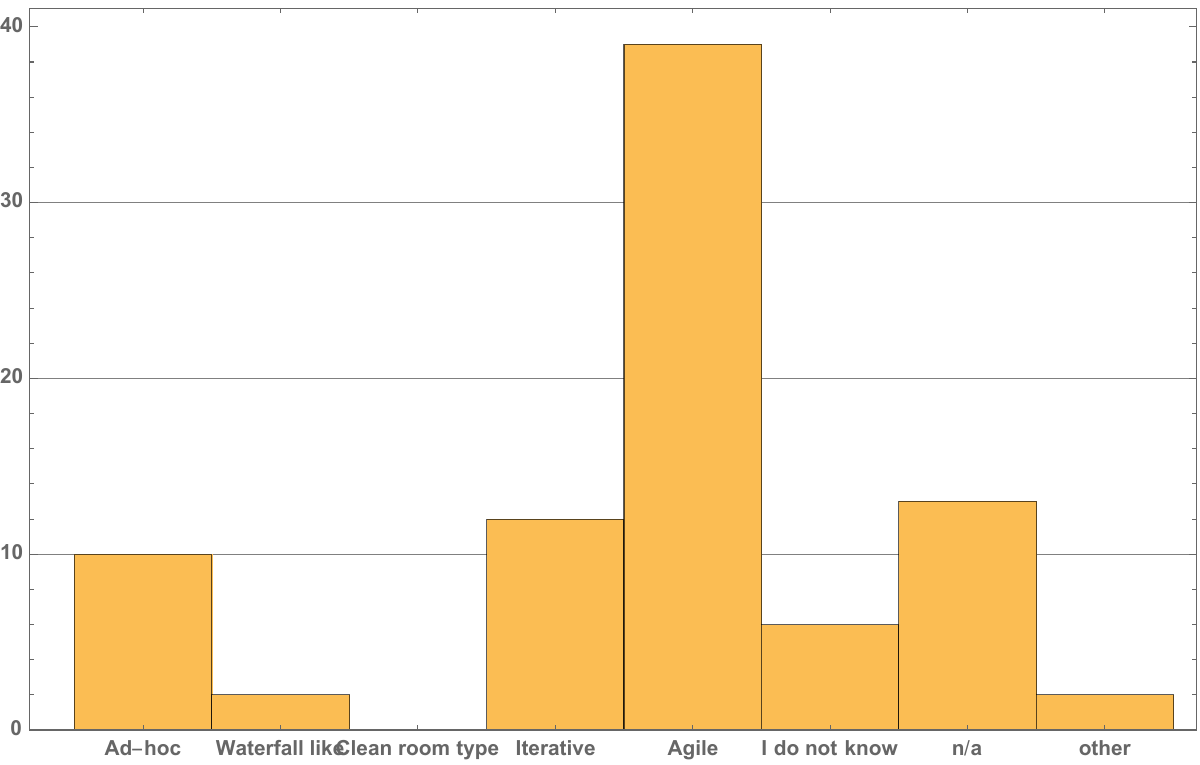}
    \caption{Use of development approaches in the studied cases}
    \label{fig_dev}
\end{figure}

\begin{figure}[ht]
    \centering
    \includegraphics[width=\textwidth]{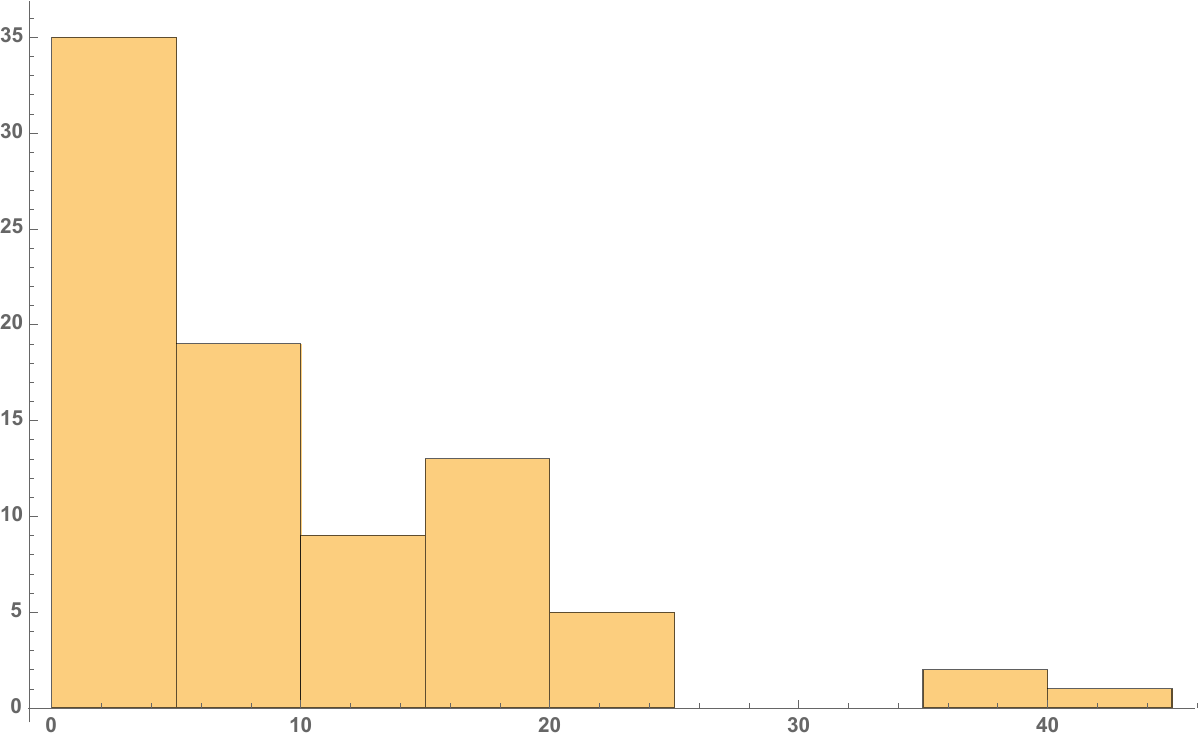}
    \caption{Use of Agile practices in the studied start-up companies}
    \label{fig_use}
\end{figure}

The responses on what development type best characterizes the company, suggest that most companies, 51 out of 84, 60\%, follow agile and iterative processes. A few, 2 out of 84, follow a waterfall like process, 10 companies report using an ad-hoc approach, see Fig.~\ref{fig_use}.

We presented respondents with a list of 56 Agile practices and asked to thick off practices that they use in their companies. 
Most start-ups use between 0 and 20 Agile practices. However, the majority of companies report using only a few practices, see Fig.~\ref{fig_use}. There is also a small cluster of companies reporting the use of more than 35 individual practices. Only 7 companies report not using any Agile practices.

\begin{figure}[ht]
    \centering
    \includegraphics[width=\textwidth]{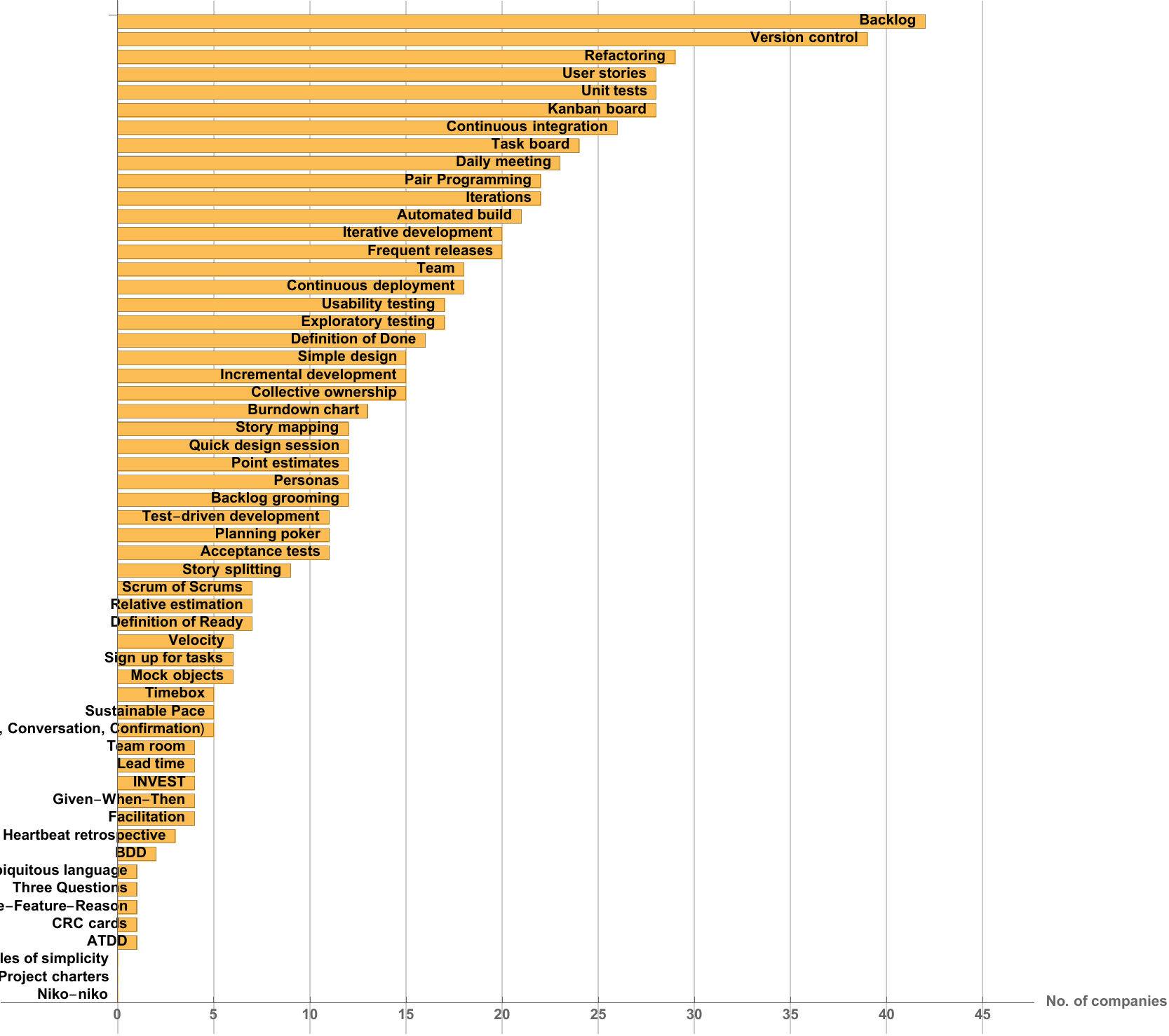}
    \caption{Frequency of Agile practices}
    \label{fig_popularity}
\end{figure}

The most frequently used Agile practices are backlog and version control reported by 42 and 39 companies, respectively (50\% and 46\% out of 84 cases). The use of other practices varies, see Fig.~\ref{fig_popularity}. Respondents do not report the use of practices such as the Niko-Niko calendar, project charters, and rules of simplicity.

\subsection{Overview of the findings}

In Table~\ref{table_agile_dimensions}, we summarize the associations between the use of certain practices and dimensions of product quality. In Table~\ref{table_agile_precedents}, we summarize the associations between the use of certain practices and dimensions of product quality. We show only practices with statistically significant associations $(p < 0.05)$. The numbers in the table show Cramers'V values denoting the strength of the associations, see Table~\ref{table_cramersv} for interpretation of the values. Up ($\uparrow$) and down ($\downarrow$) arrows denote whether the association is positive, i.e., use of the practice is associated with more positive responses, or negative, i.e., use of the practice is associated with more negative estimates from respondents.


\begin{table*}[!t]
\renewcommand{\arraystretch}{1}
\caption{Results of Cramer's V test on associations between dimensions and use of Agile practices with $p < 0.05$}
\label{table_agile_dimensions}
\centering

\begin{tabular}{L{1.7in}C{0.4in}C{0.4in}C{0.4in}C{0.4in}C{0.4in}}
Practice &  \rot{Documentation} &  \rot{Architecture} &  \rot{Source code} &  \rot{Testing} &  \rot{Overall quality}\\
\hline

Card, Conversation, Confirmation  & - & - & - & 0.422$\uparrow$ & - \\
Unit tests  & - & - & - & 0.391$\uparrow$ & -  \\
Automated build & - & - & 0.374$\uparrow$  & -  & - \\
Facilitation & - & - & 0.330$\downarrow$ &  -  & - \\
Given, When, Then & - & - & 0.330$\downarrow$  & - & -\\
INVEST & - & - & 0.330$\downarrow$ & -& -   \\
Iterations & - & 0.359$\uparrow$ & - & -& -\\
Continuous integration & - & - & - & - & 0.368$\uparrow$ \\
Collective ownership & - & - & - & - & 0.372$\downarrow$ \\
\hline

\end{tabular}
\end{table*}

\begin{table}[!h]
\renewcommand{\arraystretch}{1.1}
\caption{Results of Cramer's V test on associations between precedents of technical debt and use of Agile practices $p < 0.05$}
\label{table_agile_precedents}
\centering

\begin{tabular}{
L{1.5in}    L{0.4in}    L{0.4in}  L{0.2in}  L{0.2in}  L{0.4in}  L{0.4in} }

Practice & \rot{Attitudes} & \rot{Pragmatism}  & \rot{Communication} & \rot{Skills} & \rot{Resources} & \rot{Process}  \\
\hline

Backlog & - & - & - & - & & 0.401$\downarrow$ \\
Unit tests & 0.379$\uparrow$ & - & - & - & - & - \\
Continuous integration & 0.360$\uparrow$ & - & - & - & - & - \\
Automated build & - & - & - & - & - & 0.346$\downarrow$ \\
Definition of Done & 0.411$\downarrow$ & - & - & - & - & - \\
Simple design & - & - & - & - & 0.365$\downarrow$ & - \\
Burndown chart & 0.383$\downarrow$ & - & - & - & 0.384$\uparrow$ & - \\
Story mapping& - & 0.356$\uparrow$ & - & - & - & - \\
Relative estimation& 0.399$\downarrow$ & - & - & - & 0.399$\uparrow$ & - \\
Velocity & 0.435$\downarrow$ & - & - & - & - & - \\
Team room & - & - & - & - & 0.343$\downarrow$ & - \\
\hline

\end{tabular}
\end{table}

\subsection{Interpretations of associations}


An association shows that a specific practice and certain estimates on a factor are reported together. We use the Pearsons Chi-squared test ($p < 0.05$) to determine if the association is statistically significant. However, from associations alone, we cannot tell if there is any causal relationship between the two. Understanding of causal relationships between Agile practices, product, and team factors are crucial to guide practitioners in adopting Agile practices in start-ups. To address the causal relationships, we formulate 5 archetypes ($A$) of propositions characterizing potential explanations of our findings.



An association could point towards a causal relationship between the use of a practice ($P$) and a factor ($F$). We are measuring factors through respondents evaluations, thus we cannot distinguish between actual and perceived improvements. 
\\\\
$A_1: $ Use of $P$ improves perception of $F$.
\\\\
Some of the associations appear to be negative, i.e. use of a practice is reported together with an unfavorable estimates. It could be that the practice has adverse effects, or the use of the practice helped to expose the a problematic factor:
\\\\
$A_2: $ Use of $P$ hinders $F$.
\\\\
$A_3: $ Use of $P$ exposes issues with $F$.
\\\\
It could be that a practice is introduced as a consequence to a situation. That is, we are observing a reverse causal relationship. 
\\\\
$A_4: $ $F$ is a precedent for $P$
\\\\
It could be that a statistically significant association is a false positive. That is, the association between a practice and a factor is due to an error or some confounding factor.
\\\\
$A_0: $ There is no causal relationship between $P$ and $F$.
\\\\


\subsection{Specific findings}
In this section, we link together our specific findings with relevant propositions.

\mypara{A product backlog} is an authoritative list of new features, changes, bug fixes, and other activities that a team may deliver to achieve a specific outcome~\cite{agilealliance}.

Our results show a moderately strong $(Cramers'V = 0.401)$ association between the use of a backlog and worse perceptions on the engineering process. In particular, frequent changes in requirements, unclear objectives, and unsystematic changes hindering the engineering process are reported together with the use of the backlog. Relevant propositions: $A_0, A_2, A_3, A_4$


\mypara{Unit testing} is a practice to develop short scripts to automate examination of low-level behavior of the software~\cite{agilealliance}.

Our findings show a moderately strong association ($Cramers'V = 0.379$) between the use of unit tests and teams' attitudes. In particular, a positive attitude towards following the best design, coding, and testing practices are reported together with using unit testing. Relevant propositions: $A_0, A_1, A_4$

Our findings also show a moderately strong association ($Cramers'V = 0.391$) between the use of unit testing and less reliance on manual testing of the product.
Relevant propositions: $A_0, A_1, A_4$

\mypara{Continuous integration} aims to minimize the duration and effort of each integration episode, and maintain readiness to deliver a complete product at any moment~\cite{agilealliance}.

Our findings show a moderately strong association ($Cramers'V = 0.360$) between the use of continuous integration and more positive attitudes towards using sound design, coding, and testing practices.
Relevant propositions: $A_0, A_1, A_4$

Our findings also show a moderately strong association ($Cramers'V = 0.368$) between the use of continuous integration and more positive estimates on product internal and external quality, and less slipped defects.
Relevant propositions: $A_0, A_1, A_4$

\mypara{Automated build} is a practice to automate the steps of compiling, linking, and packaging the software for deployment~\cite{agilealliance}.

Our findings show a moderately strong ($Cramers'V = 0.346$) association between the use of automated build and worse estimates on the engineering process. 
Relevant propositions: $A_0, A_2, A_3, A_4$

Our findings also show a moderately strong ($Cramers'V = 0.374$) association between the use of automated builds and more positive estimates on the source code quality. Relevant propositions: $A_0, A_1, A_4$

\mypara{Definition of done} is a list of criteria which s task must meet before it is considered done~\cite{agilealliance}.

Our findings show a moderately strong ($Cramers'V = 0.411$) association between the use of a definition of done and worse attitudes towards following the best engineering practices.
Relevant propositions: $A_0, A_2, A_3, A_4$

\mypara{Simple design} is a practice to favor simple, modular, and reusable software designs that are created as needed~\cite{agilealliance}.

Our findings show a moderately strong association ($Cramers'V = 0.365$) between simple design practices and more pressing time and resources concerns. Relevant propositions: $A_0, A_2, A_3, A_4$

\mypara{Burndown chart} is a graph visualizing the remaining work (x-axis) over time (y-axis)~\cite{agilealliance}. 

Our findings show a moderately strong association ($Cramers'V = 0.383$) between the use of the burndown chart and worse estimates on teams' attitudes towards following the best engineering practices.
Relevant propositions: $A_0, A_2, A_3, A_4$

Our findings also show a moderately strong association ($Cramers'V = 0.384$) between the use of the burndown chart and less time and resources pressure.
Relevant propositions: $A_0, A_1, A_4$

\mypara{Story mapping} is a practice to organize user stories in a two-dimensional map according to their priority and the level of sophistication. Such a map is used to identify requirements for a bare-bones but usable first release, and subsequent levels of increased functionality~\cite{agilealliance}. 

Our findings show a moderately strong association ($Cramers'V = 0.356$) between the use of the story mapping and a more pragmatic approach on handing the trade-off between time-to-market and following the best engineering practices.
Relevant propositions: $A_0, A_1, A_4$

\mypara{Relative estimation} comprises of estimating task effort in relation to other similar tasks, and not absolute units~\cite{agilealliance}.

Our findings show a moderately strong association ($Cramers'V = 0.399$) between the use of the relative estimation and worse attitudes towards following the best testing, architecture, and coding practices.
Relevant propositions: $A_0, A_2, A_3, A_4$

Our results also show a moderately strong association ($Cramers'V = 0.399$) between the use of relative estimates and less time and resources pressure.
Relevant propositions: $A_0, A_1, A_4$

\mypara{Velocity} is a metric to calculate how long it will take to complete the project based on past performance~\cite{agilealliance}.

Our findings show a moderately strong association ($Cramers'V = 0.435$) between the use of velocity and worse attitudes towards following the best engineering practices.
Relevant propositions: $A_0, A_2, A_3, A_4$

\mypara{Team room} is a dedicated, secluded, and equipped space for an agile team to collaborate on the project~\cite{agilealliance}. 

Our findings show a moderately strong association ($Cramers'V = 0.343$) between the use of a team room and more pressing time and resource constraints.
Relevant propositions: $A_0, A_2, A_3, A_4$

\mypara{Facilitation} is a practice to have a dedicated person in the meeting, ensuring effective communication, and maintaining focus on the objectives~\cite{agilealliance}.

\mypara{Given, When, Then} is a template for formulating user stories comprising of some contextual information, triggers or actions, and a set of observable consequences~\cite{agilealliance}.

\mypara{INVEST} is a checklist to evaluate the quality of a user story~\cite{agilealliance}.

Our findings show a moderately strong association ($Cramers'V = 0.330$) between the use of any of the three practices (Facilitation, Given, When, Then, and INVEST) and worse estimates on the product source code quality.
Relevant propositions: $A_0, A_2, A_3, A_4$

\mypara{Iterations} are time-boxed intervals in an agile project in which the work is organized. A project consists of multiple iterations, tasks, and objectives for the next iteration and is revised just before it starts~\cite{agilealliance}.

Our findings show a moderately strong association ($Cramers'V = 0.359$) between the use of iterations and more positive estimates on the quality of product architecture. Specifically, respondents report fewer workarounds, more optimal selection of technologies, and fewer issues with outdated designs. 
Relevant propositions: $A_0, A_1, A_4$

\mypara{Collective ownership} is a practice to empower any developer to modify any part of the project source code~\cite{agilealliance}.

Our findings show a moderately strong association ($Cramers'V = 0.372$) between collective ownership and worse estimates on the product's internal and external quality. Relevant propositions: $A_0, A_2, A_3, A_4$

\mypara{Card, Conversation, Confirmation} is a pattern capturing the life-cycle of a user story. The life-cycle starts with tangible ``card'', ``conversations'' regarding the user story takes place throughout the project; finally, a ``confirmation'' is received of a successful implementation of the user story~\cite{agilealliance}.

Our findings show a moderately strong association ($Cramers'V = 0.422$) between the use of the life-cycle pattern and less dependence on manual testing of the product.
Relevant propositions: $A_0, A_1, A_4$

\section{Discussion}
\subsection{Answers to our research questions}

\paragraph{RQ1: How are Agile practices used in startups} 
Our results show that start-ups use Agile practices, even though they do not follow any specific agile methodology. Such results confirm earlier findings, e.g., Giardino et al.~\cite{Unterkalmsteiner}, and Yau and Murphy~\cite{Yau2013}, stating that engineering practices and processes in start-ups gradually evolve from rudimentary and ad-hoc to more systematic. 

The most frequently used practices are a backlog, version control, refactoring, user stories, unit tests, and kanban board. We could not identify any clear tendencies comparing frequencies of practices between different cohorts, e.g., team size, product stage, and team skill level. 




The use of Agile practices does not imply that an organization follows agile principles as proposed by the Agile manifesto~\cite{beck2001manifesto}. Many of the Agile practices, for example, version control, unit testing, and refactoring, among others, could be equally well applied with other types of development methodologies. That said, a majority of start-ups characterize their development methodology as agile. Exploring the maturity of agile processes in start-ups remains a direction for further exploration~\cite{gren2015prospects,klotins2019progression}.

\paragraph{RQ2: What are the associations between specific Agile practices and product factors} 
We identify associations between the use of Agile practices and product architecture, source code quality, test automation, and the overall level of quality. We could not identify any associations regarding the quality and understandability of product documentation. 

Practices related to automation, e.g., unit tests, automated build, and continuous integration, are associated with positive estimates on product factors. Practices related to requirements quality, e.g., Given, when, then, and INVEST, show negative associations. It could be that start-ups introduce such practices as a response to the adverse effects of poor requirements. However, the causal effects of using Agile practices need to be explored further to draw any definitive conclusions.

The use of collective ownership is associated with negative estimates of overall product quality. We propose two interpretations: a) collective ownership exposes the actual state of product internal quality, b) collective ownership has adverse effects. 

If two or more developers collaborate on the same part of the product, they may have a more objective view of its flaws. A single developer working on and ``owning'' a part of a product may be biased in estimating its quality~\cite{ozer2015contextualized}.

Alternatively, inviting other developers to work on the part of a product could introduce defects. Other developers, who are not the original authors, may lack the essential contextual information to evaluate and change the component without introducing defects. Practices such as unit testing, continuous integration, and pair programming may help to prevent defects and distribute knowledge in the team. 
Collective ownership could be an example of a practice that must be supported by other practices to avoid adverse effects.

\paragraph{RQ3: What are the associations between specific Agile practices and team factors}
The most associations pertain to teams' attitudes towards following the best engineering practices. Both positive and negative attitudes towards the best engineering practices are precedents for using several practices. Automation practices, such as unit tests and continuous integration, are associated with positive attitudes. However, control and planning practices, such as the definition of done, burndown chart, relative estimation, and velocity, are associated with negative attitudes towards following the best engineering practices. We explain such results with the need for tighter control over the team's performance when they do not see the benefits of following the best practices.

We observe several associations between the use of Agile practices and respondents' estimates towards time and resource pressures. The use of burndown charts and relative estimates are associated with less pressure. We interpret such findings that the use of resource planning and control practices helps to plan any amount of resources better and alleviate the pressure.

We have not identified any associations about communication in the team. Other authors, e.g., Yau et al.~\cite{Yau2013} and Sutton~\cite{Sutton2000}, have identified that in small start-up teams, communication is not an issue. Small collocated teams do not need additional support for coordination. Such finding leads us to argue that the primary reasons for introducing Agile practices in start-ups are tighter control over a team's performance and resource utilization.

\subsection{Implications to research}
With this study, we have set forth a number propositions for further investigation. Looking at the propositions, summarized in Figure X, we identify several cross-cutting concerns to address with further studies in the area.

Our results suggest that start-ups adopt Agile practices one by one without following any particular agile methodology, e.g., scrum or XP. Such finding is supported earlier work, for example, Giardino et al.~\cite{Unterkalmsteiner} and Gralha et al.~\cite{gralha2018evolution}, reporting that new practices are introduced gradually and aimed at addressing specific concerns. However, existing research on adopting agile software engineering considers mostly the adoption of whole methodologies, e.g., scrum, or XP, and not individual practices~\cite{dybaa2008empirical}. We identify an opportunity for more fine-grained research on how to adopt Agile practices in small organizations to address their specific concerns.

Related work identifies the need to be more flexible and to alleviate the need for rigorous upfront planning as the primary goal for adopting agile. Other objectives include the aim to improve product quality, shorten feedback loops with customers, and to improve teams' morale~\cite{dybaa2008empirical}. Such objectives are superficial and do not support the adoption of specific practices or addressing specific start-up specific challenges~\cite{Giardino2014}. We identify an opportunity to explore precedents of introducing specific Agile practices, and also longitudinal studies examining the effects of specific practices.



\subsection{Implications for practitioners}

Examining our findings, we identify several relevant patterns for practitioners. 

Teams' attitudes towards following the best engineering practices appear as a strong denominator of adopting a range of Agile practices. Positive attitudes towards good practices drive the adoption of automated testing and continuous integration. Such practices have further positive effects on software quality.

Negative attitudes towards the best practices are associated with the use of practices for progress control, such as the definition of done, burndown chart, and effort estimation. Our explanation for such a finding is that teams implement such practices to have tighter control over the development process.

Our results suggest that the primary benefits of adopting Agile practices are tighter control over the team's performance and product quality. The use of progress control practices alleviates resource pressures.

\section{Conclusions}
In this study, we investigate associations between the use of Agile practices and perceived impact on various product and team factors. Based on our findings, we set forth a number of propositions that narrow down the space of investigation for future studies on Agile practices and start-ups. 

We conclude that start-ups adopt Agile practices, however do not follow any specific methodology. The use of Agile practices is associated with improved product quality, more positive attitudes towards following the best engineering practices, and tighter control over resource utilization. However, the exploration of the causal effects remains a direction of further work.

We have formulated several implications for researchers and practitioners. We identify an opportunity for more fine-grained studies (on practice level) into the adoption and effects of Agile practices. We conclude that Agile practices show a potential to be used in start-ups setting, however adopting individual practices without considering the supporting practices could lead to adverse effects.








\bibliography{mybibfile}

\end{document}